\documentclass[twocolumn,showpacs,aps,amsmath,pra,amssymb,superscriptaddress]{revtex4-2}  
\usepackage[latin9]{inputenc}
\usepackage{textcomp}
\usepackage{amsmath}
\usepackage{graphicx}
\PassOptionsToPackage{version=3}{mhchem}
\usepackage{mhchem}
\usepackage{xcolor}

\makeatletter

\usepackage{subfigure}\usepackage{epstopdf}\usepackage{dcolumn}
\usepackage{bm}
\usepackage{multirow}
\usepackage{longtable}\usepackage{rotating}\usepackage{threeparttable}
\usepackage{float}
\def\br{{\bf r}}
\def\bx{{\bf x}}

\makeatother

\begin{document}

\title{Derivation of Hierarchically Correlated Orbital Functional Theory: The Role of Hypercomplex Orbitals}
\author{Ting Zhang}
\affiliation{Center for Theoretical and Computational Chemistry, Frontiers Science Center for New Organic Matter, State Key Laboratory of Advanced Chemical Power Sources, Key Laboratory of Advanced Energy Materials Chemistry (Ministry of Education), Department of Chemistry, Nankai University, Tianjin 300071, China}

\author{Neil Qiang Su}
\email[Corresponding author. ]{nqsu@nankai.edu.cn}
\affiliation{Center for Theoretical and Computational Chemistry, Frontiers Science Center for New Organic Matter, State Key Laboratory of Advanced Chemical Power Sources, Key Laboratory of Advanced Energy Materials Chemistry (Ministry of Education), Department of Chemistry, Nankai University, Tianjin 300071, China}

\begin{abstract}
This work presents a detailed mathematical derivation of the hierarchically correlated orbital functional theory (HCOFT), a framework based on hypercomplex orbitals. Recent study [Phys. Rev. Lett. 133, 206402] has demonstrated that hypercomplex orbitals in a determinant are equivalent to a set of real-valued orbitals that allow fractional occupations, making them desirable fundamental descriptors for many-electron systems. The algebraic properties of Clifford algebra are rigorously applied to derive key quantities within HCOFT, addressing the complexities introduced by the hypercomplex representation. It is shown that, despite this added complexity, the resulting density and kinetic energy remain physically meaningful and satisfy essential properties, including the Pauli exclusion principle. To establish the uniqueness of HCOFT, alternative definitions of hypercomplex orbitals within Clifford algebra are explored. These alternatives can lead to the loss of physical meaning in fundamental quantities for many-electron systems. Overall, this work demonstrates that HCOFT not only preserves the desired physical properties but also provides a single-determinant framework capable of describing multi-reference systems.
\end{abstract}
\maketitle

\section{\label{sec:int} Introduction}

Kohn-Sham density functional theory (KS-DFT) \cite{HK1964,Levy1979pnas,KS1965,PY1989,Dreizler2012} has proven to be a powerful tool for electronic structure calculations, particularly in systems where a single-determinant description is sufficient. However, it faces challenges when dealing with strong correlation effects, as traditional functionals in this framework systematically fall short in multi-reference systems \cite{Cohen2008science,Paula2009prl,Burke2012jcp,Cohen2012cr,Becke2014jcp,Mardirossian2017,Xu2004,DFT2022PCCP}. This limitation necessitates extensions beyond the traditional single-determinant framework. One promising approach is reduced density matrix functional theory (RDMFT) \cite{Gilbert1975prb,Levy1979pnas,Valone1980jcp}, which incorporates the full density matrix to capture more subtle correlation effects. However, RDMFT still requires further improvements in both efficiency and accuracy, and its development is ongoing \cite{Muller1984rpa,GU1998prl,Pernal2005prl,Sharma2008prb,Piris2010jcp,Sharma2013prl,Schade2017,Christian2018,Schilling2019prl,Cioslowski2020jctc,Cioslowski2020jcp,Yao2021jpcl,Piris2021prl,Gibney2021jpcl,Yao2022jpca,Gibney2022jctc,Ai2022jpcl,Ai2023jcp,Gibney2023prl,Liebert2023jcp,Yao2024jpca,Cartier2024jctc,Cartier2025arXiv,Voutou2025arXiv}.

The introduction of hypercomplex orbitals provides a rigorous pathway for transitioning from KS-DFT to RDMFT \cite{Su2021pra}. Hypercomplex orbitals offer additional degrees of freedom while maintaining essential quantum mechanical properties, enabling a more versatile and precise description of correlated electronic states. This feature facilitates a distinct formulation of the functional theory that, while staying within the single-determinant formalism, overcomes the traditional limitations of the single-determinant approximation. Furthermore, recent study has demonstrated that hypercomplex orbitals with integer occupations are equivalent to a set of real orbitals that allow fractional occupations, referred to as hierarchically correlated orbitals (HCOs), when considered as fundamental descriptors of the system \cite{Su2024prl}. This finding is particularly intriguing, as it highlights two distinct perspectives for describing electronic systems: one using integer-occupied hypercomplex orbitals and the other employing fractionally occupied real orbitals.

To enhance the understanding of key concepts and foster the development of advanced methods, this work provides a formal derivation of the functional theory using hypercomplex orbitals. The main objectives of this study are twofold: first, to offer a comprehensive presentation of the theoretical foundations of HCO functional theory (HCOFT) (Sec. \ref{sec2}) and derive key expressions for fundamental quantities within this framework (Sec. \ref{sec3}); second, to explore new extensions of the hypercomplex orbital representation that may further improve the functional's ability to describe complex correlated systems (Sec. \ref{sec4}). These advancements aim to contribute to the development of HCOFT as an alternative framework that systematically incorporates strong correlation effects, extending beyond the limitations of the traditional single-determinant formalism found in KS-DFT.


\section{\label{sec2} Functional Theory using HyperComplex Orbitals}

This section provides a detailed review of the fundamental concepts in HCOFT, along with a more comprehensive derivation to elaborate the previous work \cite{Su2024prl}. Unless otherwise specified, the indices \(i\), \(j\), and \(k\) are used to index occupied orbitals, while \(p\) and \(q\) index all orbitals.

In a single-determinant formalism, the $N$-electron wavefunction is represented by a Slater determinant \cite{Szabo1996modern}:
\begin{eqnarray}
|\Phi\rangle=\frac{1}{\sqrt{N!}}\left|\begin{array}{cccc}
    \psi_1({\bx}_1) & \psi_2(\bx_1) & \cdots & \psi_N(\bx_1) \\
    \psi_1(\bx_2) & \psi_2(\bx_2) & \cdots & \psi_N(\bx_2) \\
    \vdots                        & \vdots                        & \vdots & \vdots  \\
    \psi_1(\bx_N) & \psi_2(\bx_N) & \cdots & \psi_N(\bx_N) 
\end{array}\right| \label{eq:wf_ks}  \\
=\!\frac{1}{\sqrt{N!}}\!\sum_{(i_1i_2\cdots i_N)}\!(\!-\!1\!)^{\tau(i_1i_2\cdots i_N)}\!\psi_{i_1}\!(\!\bx_1\!)\!\psi_{i_2}\!(\!\bx_2\!)\!\cdots\!\psi_{i_N}\!(\!\bx_N\!), \label{eq:dete}
\end{eqnarray}
where the summation extends over all $N!$ permutations of $(12 \cdots N)$, and $\tau(i_1 i_2 \cdots i_N)$ denotes the inversion number of the permutation $(i_1 i_2 \cdots i_N)$. The variable $\bx$ represents the combined spatial and spin coordinate, $(\br, \omega)$.

The spin-orbitals are chosen as eigenstates of the spin projection along the $z$-axis, factorized as \cite{Szabo1996modern}
\begin{equation}
\label{eq:orbi}
\psi_{i}(\bx)=\varphi_{\tilde{i}}^{\sigma_{i}}(\br)\sigma_{i}(\omega).
\end{equation}
For a system with $N_\alpha$ spin-up ($\alpha$) and $N_\beta$ spin-down ($\beta$) electrons, such that $N = N_\alpha + N_\beta$, the spin states are
\begin{eqnarray}
\label{eq:sigma}
\sigma_{i} =
\begin{cases}
    \alpha, & 1 \leq {i} \leq N_\alpha \\
    \beta, & N_\alpha < {i} \leq N  
\end{cases}
\end{eqnarray}
which obey the orthonormality conditions:
\begin{equation}
\label{eq:spinortho}
\langle\alpha|\alpha\rangle=\langle\beta|\beta\rangle=1, \langle\alpha|\beta\rangle=0.
\end{equation}
Thus, the spin-orbitals $\{\psi_i\}$ naturally partition into two subsets corresponding to $\alpha$ and $\beta$ spins. 
To ensure that the occupied spatial orbitals $\{\varphi_{\tilde{i}}^{\sigma_{i}}\}$ are properly indexed from 1 within each spin sector, the index $\tilde{i}$ is defined as follows:
\begin{eqnarray}
\label{eq:sigma}
\tilde{i} =
\begin{cases}
    i, & 1 \leq i \leq N_\alpha \\
    i-N_\alpha, & N_\alpha < i \leq N 
\end{cases}
\end{eqnarray}
All spatial orbitals $\{\varphi_{p}^{\sigma}\}$ form a set of $K$ orthonormal one-electron functions,
\begin{equation}
\label{eq:ortho}
\langle \varphi_{p}^{\sigma} | \varphi_q^\sigma \rangle=\delta_{pq},
\end{equation}
spanning a space of dimension $K$ and satisfying the completeness relation
\begin{equation}
\label{eq:iden}
\sum_{p}|\varphi_{p}^{\sigma}\rangle\langle\varphi_{p}^{\sigma}|=\hat{I}_{K},
\end{equation}
where $\delta_{pq}$ is the Kronecker delta function and $\hat{I}_K$ is the identity operator in the $K$-dimensional space of the orbitals.

In HCOFT, a basis of dimension $n$ in a Clifford algebra, denoted as $\{e_1, e_2, \dots, e_n\}$, is introduced \cite{HC1989,CA2019}, satisfying the algebraic properties:
\begin{equation}
\label{eq:CA}
e_\mu^2=-1; e_\mu e_\nu=-e_\nu e_\mu.
\end{equation}
This structure enables the definition of hypercomplex orbitals as \cite{Su2021pra}
\begin{equation}
\label{eq:orb_hc}
\varphi_p^\sigma(\br)=\phi_{p}^{\sigma\!,0}(\br)+\sum_{\mu=1}^{n}\phi_{p}^{\sigma\!,\mu}(\br) e_\mu, 
\end{equation}
where $\{\phi_{p}^{\sigma,\mu}\}$ are real-valued functions. The Clifford conjugate of Eq.~\eqref{eq:orb_hc} is given by \cite{Su2021pra,CA2019}:
\begin{equation}
\label{eq:orb_conj}
\overline{\varphi}_p^\sigma(\br)=\phi_{p}^{\sigma\!,0}(\br)-\sum_{\mu=1}^{n}\phi_{p}^{\sigma\!,\mu}(\br) e_\mu.
\end{equation}

Substituting Eqs.~\eqref{eq:orb_hc} and \eqref{eq:orb_conj} into Eq.~\eqref{eq:ortho}, the orthonormality condition transforms into:
\begin{align}
\label{eq:ortho1}
& \sum_{\mu=0}^n \langle\phi_p^{\sigma,\mu}|\phi_q^{\sigma,\mu}\rangle+\sum_{\mu=1}^n (\langle\phi_p^{\sigma,0}|\phi_q^{\sigma,\mu}\rangle-\langle\phi_p^{\sigma,\mu}|\phi_q^{\sigma,0}\rangle)e_\mu \nonumber \\
&+\sum_{1\leq\mu<\nu\leq n} (\langle\phi_p^{\sigma,\nu}|\phi_q^{\sigma,\mu}\rangle-\langle\phi_p^{\sigma,\mu}|\phi_q^{\sigma,\nu}\rangle)e_\mu e_\nu=\delta_{pq},
\end{align}
This condition implies:
\begin{align}
\label{eq:ortho2}
\begin{cases}
\sum_{\mu=0}^n \langle\phi_p^{\sigma,\mu}|\phi_q^{\sigma,\mu}\rangle=\delta_{pq},  \\
\langle\phi_p^{\sigma,\mu}|\phi_q^{\sigma,\nu}\rangle=\langle\phi_p^{\sigma,\nu}|\phi_q^{\sigma,\mu}\rangle,  
\end{cases}
\end{align}

Similarly, inserting Eq.~\eqref{eq:orb_hc} and \eqref{eq:orb_conj} into Eq.~\eqref{eq:iden} leads to:
\begin{align}
\label{eq:iden1}
& \sum_{\mu=0}^n\sum_p |\phi_p^{\sigma,\mu}\rangle\langle\phi_p^{\sigma,\mu}| \nonumber \\
&+\sum_{\mu=1}^n\sum_p (|\phi_p^{\sigma,\mu}\rangle\langle\phi_p^{\sigma,0}|-|\phi_p^{\sigma,0}\rangle\langle\phi_p^{\sigma,\mu}|)e_\mu \nonumber \\
&+\sum_{1\leq\mu<\nu\leq n}\sum_p (|\phi_p^{\sigma,\mu}\rangle\langle\phi_p^{\sigma,\nu}|-|\phi_p^{\sigma,\nu}\rangle\langle\phi_p^{\sigma,\mu}|)e_\mu e_\nu \nonumber \\
&=\hat{I}_K.
\end{align}
From this, the completeness relation simplifies to:
\begin{align}
\label{eq:iden2}
\begin{cases}
\sum_{\mu=0}^n \sum_p |\phi_p^{\sigma,\mu}\rangle\langle\phi_p^{\sigma,\mu}|=\hat{I}_K,  \\
\sum_p|\phi_p^{\sigma,\mu}\rangle\langle\phi_p^{\sigma,\nu}|=\sum_p|\phi_p^{\sigma,\nu}\rangle\langle\phi_p^{\sigma,\mu}|,  
\end{cases}
\end{align}
 
For constructing $|\Phi\rangle$, $N_\alpha$ and $N_\beta$ hypercomplex orbitals corresponding to two spins are required. Several methods exist for computing the determinant in Eq.~\eqref{eq:wf_ks} within hypercomplex number systems (commonly referred to as the hyperdeterminant) \cite{Cayley1845,Study1920,Moore1922,Chen1991,Kyrchei2008,Aslaksen1996}. In this work, we continue to use Eq.\eqref{eq:dete}, which serves as a direct extension of the determinant from the real and complex cases to the hypercomplex case. Additionally, since the determinant vanishes if two orbitals in $\{\psi_i\}$ are identical, Eq.~\eqref{eq:wf_ks} inherently satisfies the Pauli principle.  

According to Ref.~\cite{Lounesto2001,Su2021pra}, the Clifford conjugate reverses the order of a product of hypercomplex orbitals:
\begin{equation}
\label{eq:prod_conj}
\overline{\varphi_{\tilde{i}_1}^{\sigma_{i_1}}\!(\!\br_1\!)\varphi_{\tilde{i}_2}^{\sigma_{i_2}}\!(\!\br_2\!)\cdots\varphi_{\tilde{i}_{N}}^{\sigma_{i_N}}\!(\!\br_{\!N}\!)}=\overline{\varphi}_{\tilde{i}_N}^{\sigma_{i_N}}\!(\!\br_{\!N}\!)\!\cdots\!\overline{\varphi}_{\tilde{i}_2}^{\sigma_{i_2}}\!(\!\br_2\!)\!\overline{\varphi}_{\tilde{i}_1}^{\sigma_{i_1}}\!(\!\br_1\!),
\end{equation}
Thus, the left determinant form of Eq.~\eqref{eq:wf_ks} is expressed as:
\begin{equation}
\label{eq:dete_l}
\langle\Phi|=\!\frac{1}{\sqrt{N!}}\!\sum_{(i_1i_2\cdots i_N)}\!(\!-\!1\!)^{\tau(i_1i_2\cdots i_N)}\!\overline{\psi}_{i_N}\!(\!\bx_N\!)\!\cdots\!\overline{\psi}_{i_2}\!(\!\bx_2\!)\!\overline{\psi}_{i_1}\!(\!\bx_1\!)
\end{equation}

With this formulation, the electron density is obtained as:
\begin{equation}
\label{eq:density}
\rho_n(\br)=\langle \Phi | \hat{\rho}(\br)|\Phi\rangle=\rho^\alpha(\br) + \rho^\beta(\br),
\end{equation}
where the density operator is $\hat{\rho}(\br)=\sum_{k=1}^N\delta(\br-\br_k)$,  the densities of both spins are
\begin{equation}
\label{eq:spinden}
\rho^\sigma(\br)=\sum_{i=1}^{N_\sigma}\overline{\varphi}_{i}^{\sigma}(\br)\varphi_{i}^{\sigma}(\br),
\end{equation}
and the kinetic energy is given by:
\begin{equation}
\label{eq:Tn}
T_n=\langle \Phi | \hat{T}|\Phi\rangle=-\frac{1}{2}\sum_\sigma\sum_{i=1}^{N_\sigma}\langle\varphi_{i}^{\sigma}| \nabla^2|\varphi_{i}^{\sigma}\rangle.
\end{equation}
Unlike the conventional case involving real or complex orbitals, the derivations of Eqs.~\eqref{eq:density}-\eqref{eq:Tn} involve additional complexity, which will be detailed in the following section.

The set $\{\phi_{p}^{\sigma,\mu}\}$ can be expanded in terms of a basis set. Without loss of generality, each function is expressed as a linear combination of a set of orthonormal functions $\{\xi_p\}$
\begin{equation} 
\label{eq:phi}
\phi_p^{\sigma,\mu}(\textbf{r}) = \sum_q \xi_q(\textbf{r}) V_{pq}^{\sigma,\mu},
\end{equation}
where $\{{V}^{\sigma,\mu}\}$ represents a set of $n+1$ $K \times K$ matrices, with each ${V}^{\sigma,\mu}$ containing the expansion coefficients for the $\mu$-th component of the hypercomplex orbitals.

With this formulation, Eqs.~\eqref{eq:ortho2} and \eqref{eq:iden2} can be rewritten in terms of $\{{V}^{\sigma,\mu}\}$ as \cite{Su2021pra}
\begin{align}
\label{eq:cond_hc1}
\begin{cases}
\sum_{\mu=0}^{n} {V}^{\sigma,\mu}{V}^{\sigma,\mu T} = {I}_K,  \\
{V}^{\sigma,\mu}{V}^{\sigma,\nu T} = {V}^{\sigma,\nu}{V}^{\sigma,\mu T}, 
\end{cases}
\end{align}
and
\begin{align}
\label{eq:cond_hc2}
\begin{cases}
\sum_{\mu=0}^{n} {V}^{\sigma,\mu T}{V}^{\sigma,\mu} = {I}_K,  \\
{V}^{\sigma,\mu T}{V}^{\sigma,\nu} = {V}^{\sigma,\nu T}{V}^{\sigma,\mu},
\end{cases}
\end{align}
where ${I}_K$ is the $K \times K$ identity matrix, and the superscript $T$ denotes the matrix transpose.

Further derivation reformulates the electron density in Eq.~\eqref{eq:spinden} as \cite{Su2021pra}:
\begin{equation}
\label{eq:rho_hc}
\rho^{\sigma}_{\{\chi_p^\sigma, \lambda_p^\sigma\}}(\br) = \sum_i^{N_\sigma} \sum_{\mu=0}^n |\phi_{i}^{\sigma,\mu}(\mathbf{r})|^2 = \sum_{p=1}^{K}{\lambda}_{p}^{\sigma} |\chi_p^{\sigma}(\br)|^2.
\end{equation}
Here, $\{\chi_p^\sigma, \lambda_p^\sigma\}$ correspond to the eigenvectors and eigenvalues of the matrix $D^\sigma$:
\begin{equation}
\label{eq:D_hc}
D^\sigma = \sum_{\mu=0}^{n} {V}^{\sigma,\mu T} {\rm{I}}^{N_\sigma}_K {V}^{\sigma,\mu},
\end{equation}
where ${\rm{I}}^{N_\sigma}_K$ is a $K \times K$ diagonal matrix with the first $N_\sigma$ diagonal entries set to 1 and the remaining entries set to 0. The matrix $D^\sigma$ is then diagonalized as:
\begin{equation}
\label{eq:D_de}
D^\sigma = U^\sigma \Lambda^\sigma U^{\sigma T},
\end{equation}
where $\Lambda^\sigma$ is a diagonal matrix containing the real eigenvalues $\{\lambda_{p}^\sigma\}$, and $U^\sigma$ is a unitary matrix. The eigenvectors $\{\chi_{p}^\sigma\}$ are defined as \(\chi_{p}^\sigma = \sum_{q} \xi_q U_{qp}^\sigma\), ensuring that they remain orthonormal:
\begin{equation}
\label{eq:cons_hco}
\langle \chi_{p}^\sigma | \chi_{q}^\sigma \rangle = \delta_{pq}.
\end{equation}

The kinetic energy $T_n$ from Eq.~\eqref{eq:Tn} can be reformulated in terms of $\{\chi_p^\sigma, \lambda_p^\sigma\}$ as \cite{Su2021pra}:
\begin{align}
\label{eq:T_hc}
T_n[\{\chi_p^\sigma, \lambda_p^\sigma\}]&=-\frac{1}{2}\sum_{\sigma}\sum_{i}^{N_\sigma}\sum_{\mu=0}^n  \langle \phi_i^{\sigma,\mu} | \nabla^2 | \phi_i^{\sigma,\mu} \rangle \nonumber \\
&=-\frac{1}{2}\sum_{\sigma}\sum_{p} \lambda_{p}^{\sigma} \langle \chi_p^\sigma | \nabla^2 | \chi_p^{\sigma} \rangle.
\end{align}

Here, $\{\chi_p^\sigma, \lambda_p^\sigma\}$ are the so-called HCOs and their corresponding occupations \cite{Su2022es}. It has been demonstrated that $\{\chi_p^\sigma, \lambda_p^\sigma\}$ can serve as the fundamental descriptors of the system, thereby leading to the formulation of HCOFT \cite{Su2024prl}. The ground-state energy is obtained by minimizing the following energy functional:
\begin{align}
\label{eq:totalene}
E&_{n,v}^{\rm hyper}[\{\chi_p^\sigma, \lambda_p^\sigma\}]\nonumber \\
& =T_n[\{\chi_p^\sigma, \lambda_p^\sigma\}]+J[\rho^{\sigma}_{\{\chi_p^\sigma, \lambda_p^\sigma\}}] \nonumber \\
&+E_{n,xc}[\{\chi_p^\sigma, \lambda_p^\sigma\}]+\sum_\sigma \int v(\br) \rho^{\sigma}_{\{\chi_p^\sigma, \lambda_p^\sigma\}}(\br)d\br,
\end{align}
which includes the kinetic energy, Coulomb energy, exchange-correlation energy, and external potential energy. The dimension $n$ of the basis $\{e_1, e_2, \dots, e_n\}$ determines the constraints imposed on $\{\chi_p^\sigma, \lambda_p^\sigma\}$\cite{Su2021pra,Zhang2023pra,Su2024prl}, where a larger $n$ allows for the exploration of a broader domain of densities \cite{Su2024prl}. Additionally, all occupations of the HCOs are constrained within the interval $[0,1]$, ensuring the satisfaction of the Pauli exclusion principle. Furthermore, by dynamically varying fractional occupations, the multireference nature of the system can be captured \cite{Su2021pra}, making HCOFT well-suited for describing strongly correlated systems.

\section{\label{sec3} Derivation of Fundamental Quantities}

This section provides the detailed derivations of Eqs.~\eqref{eq:density}-\eqref{eq:Tn}. By substituting Eqs.~\eqref{eq:dete} and \eqref{eq:dete_l} into Eq.~\eqref{eq:density}, one obtains:
\begin{align}
\label{eq:phirhophi}
&\langle\Phi|\hat{\rho}(\br)|\Phi\rangle\nonumber \\
&=\frac{1}{N!}\!\sum_{k=1}^N\!\sum_{(i_1i_2\cdots i_N)}\!\sum_{(i'_1i'_2\cdots i'_N)}\!(\!-\!1\!)^{\tau(i_1i_2\cdots i_N)+\tau(i'_1i'_2\cdots i'_N)} \nonumber \\
&\ \ \ \ \int d\bx_N\cdots\int d\bx_2\int d\bx_1
\!\overline{\psi_{i_1}\!(\!\bx_1\!)\!\psi_{i_2}\!(\!\bx_2\!)\!\cdots\!\psi_{i_N}\!(\!\bx_N\!)}  \nonumber \\
&\ \ \ \ \ \ \delta(\br-\br_k)\!\psi_{i'_1}\!(\!\bx_1\!)\!\psi_{i'_2}\!(\!\bx_2\!)\!\cdots\!\psi_{i'_N}\!(\!\bx_N\!).
\end{align}
Further inserting Eq.~\eqref{eq:orbi}, the multiple integrals split into two parts:
\begin{align}
\label{eq:phirhophi_2}
&\langle\Phi|\hat{\rho}(\br)|\Phi\rangle\nonumber \\
&=\frac{1}{N!}\!\sum_{k=1}^N\!\sum_{(i_1i_2\cdots i_N)}\!\sum_{(i'_1i'_2\cdots i'_N)}\!(\!-\!1\!)^{\tau(i_1i_2\cdots i_N)+\tau(i'_1i'_2\cdots i'_N)} \nonumber \\
&\ \ \ \ \mathcal{S}_{(i_1i_2\cdots i_N)}^{(i'_1i'_2\cdots i'_N)} \cdot \mathcal{O}_{(i_1i_2\cdots i_N),k}^{(i'_1i'_2\cdots i'_N)}\!(\br).
\end{align}
The spin part is given by: 
\begin{equation}
\label{eq:spinint}
\mathcal{S}_{(i_1i_2\cdots i_N)}^{(i'_1i'_2\cdots i'_N)}=\prod_{j=1}^N\langle \sigma_{i_j}|\sigma_{i'_j}\rangle,
\end{equation}
which vanishes unless each spin-orbital pair $(\psi_{i_j},\psi_{i'_j})$ corresponds to the same spin, as ensured by Eq.~\eqref{eq:spinortho}.  
The spatial part is given by:
\begin{align}
\label{eq:spatialint}
&\mathcal{O}_{(i_1i_2\cdots i_N),k}^{(i'_1i'_2\cdots i'_N)}\!(\br) \nonumber \\
&=\int d\br_{\!N}\cdots\int d\br_2\int d\br_1
\!\overline{\varphi}_{\tilde{i}_N}^{\sigma_{i_N}}\!(\!\br_{\!N}\!)\!\cdots\!\overline{\varphi}_{\tilde{i}_2}^{\sigma_{i_2}}\!(\!\br_2\!)\!\overline{\varphi}_{\tilde{i}_1}^{\sigma_{i_1}}\!(\!\br_1\!) \nonumber \\
&\ \ \ \ \ \ \delta(\br-\br_k)\!\varphi_{\tilde{i}'_{1}}^{\sigma_{i'_1}}\!(\!\br_1\!)\!\varphi_{\tilde{i}'_2}^{\sigma_{i'_2}}\!(\!\br_2\!)\!\cdots\!\varphi_{\tilde{i}'_N}^{\sigma_{i'_N}}\!(\!\br_{\!N}\!).
\end{align}
Unlike real or complex orbitals, exchanging two hypercomplex orbitals in Eq.~\eqref{eq:spatialint} alters the expression, making the multiple integrals intractable. The following section presents a lemma that elegantly bypasses these integrals.

Due to Eq.~\eqref{eq:spinint}, the integrals in Eq.~\eqref{eq:spatialint} only need to consider cases where each spin-orbital pair $(\psi_{i_j},\psi_{i'_j})$ has the same spin for all $j$. Utilizing the orthonormality condition in Eq.~\eqref{eq:ortho}, Eq.~\eqref{eq:spatialint} simplifies to:
\begin{align}
\label{eq:spatialint2}
&\mathcal{O}_{(i_1i_2\cdots i_N),k}^{(i'_1i'_2\cdots i'_N)}\!(\br) \nonumber \\
=\Big\{&\int \!\overline{\varphi}_{\tilde{i}_N}^{\sigma_{i_N}}\!(\!\br_{\!N}\!)\cdots \Big[\!\int\!\overline{\varphi}_{\tilde{i}_{k+1}}^{\sigma_{i_{k+1}}}\!(\!\br_{k+1}\!)\nonumber \\
&\Big(\!\int\!\overline{\varphi}_{\tilde{i}_k}^{\sigma_{i_k}}\!(\!\br_k\!)\!\delta(\br-\br_k)\!\varphi_{\tilde{i}'_{k}}^{\sigma_{i'_k}}\!(\!\br_k\!)d\br_k\!\Big)\nonumber \\
&\varphi_{\tilde{i}'_{k+1}}^{\sigma_{i'_{k+1}}}\!(\!\br_{k+1}\!)d\br_{k+1}\!\Big] \!\cdots\!\varphi_{\tilde{i}'_N}^{\sigma_{i'_N}}\!(\!\br_{\!N}\!)d\br_{\!N}\Big\}\prod_{j=1}^{k-1}\delta_{i_ji'_j}.
\end{align}
For the integral inside the parentheses, inserting Eqs.~\eqref{eq:orb_hc} and \eqref{eq:orb_conj} gives:
\begin{align}
\label{eq:int_k}
&\int\!\overline{\varphi}_{\tilde{i}_k}^{\sigma_{i_k}}\!(\!\br_k\!)\!\delta(\br-\br_k)\!\varphi_{\tilde{i}'_{k}}^{\sigma_{i'_k}}\!(\!\br_k\!)d\br_k\nonumber \\
&=\overline{\varphi}_{\tilde{i}_k}^{\sigma_{i_k}}\!(\!\br\!)\!\varphi_{\tilde{i}'_{k}}^{\sigma_{i'_k}}\!(\!\br\!) \nonumber \\
&=\mathcal{A}_{i_k}^{i'_k}\!(\br)\!+\!\sum_{\mu=1}^n\!\mathcal{B}_{i_k,\mu}^{i'_k}\!(\br)e_\mu\!+\!\sum_{1\leq\mu<\nu\leq n}\!\mathcal{C}_{i_k,\mu\nu}^{i'_k}\!(\br)e_\mu e_\nu,
\end{align}
where 
\begin{equation}
\label{eq:Ak}
\mathcal{A}_{i_k}^{i'_k}\!(\br)=\sum_{\mu=0}^n \phi_{\tilde{i}_k}^{\sigma_{i_k},\mu}\!(\br)\phi_{\tilde{i}'_k}^{\sigma_{i'_k},\mu}\!(\br),
\end{equation}
\begin{equation}
\label{eq:Bk}
\mathcal{B}_{i_k,\mu}^{i'_k}\!(\br)= \phi_{\tilde{i}_k}^{\sigma_{i_k},0}\!(\br)\phi_{\tilde{i}'_k}^{\sigma_{i'_k},\mu}\!(\br)- \phi_{\tilde{i}_k}^{\sigma_{i_k},\mu}\!(\br)\phi_{\tilde{i}'_k}^{\sigma_{i'_k},0}\!(\br),
\end{equation}
and
\begin{equation}
\label{eq:Ck}
\mathcal{C}_{i_k,\mu\nu}^{i'_k}\!(\br)= \phi_{\tilde{i}_k}^{\sigma_{i_k},\nu}\!(\br)\phi_{\tilde{i}'_k}^{\sigma_{i'_k},\mu}\!(\br)- \phi_{\tilde{i}_k}^{\sigma_{i_k},\mu}\!(\br)\phi_{\tilde{i}'_k}^{\sigma_{i'_k},\nu}\!(\br),
\end{equation}
which satisfies the antisymmetry property: $\mathcal{C}_{i_k,\mu\nu}^{i'_k}\!(\br)=-\mathcal{C}_{i_k,\nu\mu}^{i'_k}\!(\br)$. 

Notably, when $i_k = i'_k$, then $\mathcal{B}_{i_k,\mu}^{i_k}(\br) = \mathcal{C}_{i_k,\mu\nu}^{i_k}(\br) = 0$, reducing Eq.~\eqref{eq:int_k} to a real function that commutes with any hypercomplex orbital. This results in:
\begin{equation}
\label{eq:spatialint3}
\mathcal{O}_{(i_1i_2\cdots i_N),k}^{(i'_1i'_2\cdots i'_N)}\!(\br) =\mathcal{O}_{(i_1i_2\cdots i_N),k}^{(i_1i_2\cdots i_N)}\!(\br)\prod_{j=1}^{N}\delta_{i_ji'_j},
\end{equation}
Thus, only the terms where $(i_1 i_2 \cdots i_N) = (i'_1 i'_2 \cdots i'_N)$ survive in this case, yielding:
\begin{equation}
\label{eq:spatialint4}
\mathcal{O}_{(i_1i_2\cdots i_N),k}^{(i_1i_2\cdots i_N)}\!(\br)=\overline{\varphi}_{\tilde{i}_k}^{\sigma_{i_k}}(\br)\varphi_{\tilde{i}_k}^{\sigma_{i_k}}(\br)
=\mathcal{A}_{i_k}^{i_k}\!(\br).
\end{equation}

For cases where $i_k \neq i'_k$ or for general cases where $(i_1 i_2 \cdots i_N) \neq (i'_1 i'_2 \cdots i'_N)$, the multiple spatial integrals in Eq.~\eqref{eq:spatialint2} can be handled by using the following lemma:

\textbf{Lemma 1:}  
Let $X$ be hypercomplex, of the form:
\begin{equation}
\label{eq:X}
X = A + \sum_{\mu=1}^n B_\mu e_\mu + \sum_{1\leq\mu<\nu\leq n} C_{\mu\nu} e_\mu e_\nu,
\end{equation}
where $A$, $B_\mu$, and $C_{\mu\nu}$ are real valued, and $C_{\mu\nu} = -C_{\nu\mu}$. Then, the bra-ket integral over two hypercomplex spatial orbitals, that is $\langle \varphi_p^\sigma | X | \varphi_q^\sigma \rangle$, retains the same hypercomplex structure:
\begin{equation}
\label{eq:Xp}
\langle \varphi_p^\sigma | X | \varphi_q^\sigma \rangle = A' + \sum_{\mu=1}^n B'_\mu e_\mu + \sum_{1\leq\mu<\nu\leq n} C'_{\mu\nu} e_\mu e_\nu,
\end{equation}
where $A'$, $B'_\mu$, and $C'_{\mu\nu}$ are real valued satisfying the antisymmetry condition $C'_{\mu\nu} = -C'_{\nu\mu}$. Moreover, the scalar component transforms as:
\begin{equation}
A' = \delta_{pq} A.
\end{equation}

\textbf{Proof 1:} By inserting Eq.~\eqref{eq:X} into Eq.~\eqref{eq:Xp}, the following expression is obtained:
\begin{align}
\label{eq:varphiXvarphi}
\langle\varphi_p^\sigma|X|\varphi_q^\sigma\rangle&=\langle\varphi_p^\sigma|A|\varphi_q^\sigma\rangle \nonumber \\
&+\langle\varphi_p^\sigma|\sum_{\mu=1}^n B_\mu e_\mu|\varphi_q^\sigma\rangle \nonumber \\
&+\langle\varphi_p^\sigma|\sum_{1\leq\mu<\nu\leq n} C_{\mu\nu}e_\mu e_\nu|\varphi_q^\sigma\rangle,
\end{align}
The relations in Eq.~\eqref{eq:CA} are used to evaluate the three terms on the right-hand side of Eq.~\eqref{eq:varphiXvarphi}.
The first term is straightforward:
\begin{equation}
\label{eq:varphiXvarphi-1}
\langle\varphi_p^\sigma|A|\varphi_q^\sigma\rangle=\delta_{pq}A.
\end{equation}
For the second term, first consider:
\begin{align}
\label{eq:varphiX-2}
&\overline{\varphi}_p^\sigma(\br)\sum_{\mu=1}^n B_\mu e_\mu =\Big[\phi_{p}^{\sigma\!,0}(\br)-\sum_{\nu=1}^{n}\phi_{p}^{\sigma\!,\nu}(\br) e_\nu\Big] \sum_{\mu=1}^n B_\mu e_\mu \nonumber \\
&=-\sum_{\mu=1}^n B_\mu e_\mu \overline{\varphi}_p^\sigma(\br)+2\sum_{\mu=1}^n B_\mu \phi_{p}^{\sigma\!,\mu}(\br)+2\sum_{\mu=1}^n B_\mu \phi_{p}^{\sigma\!,0}(\br) e_\mu.
\end{align}
This leads to:
\begin{align}
\label{eq:varphiXvarphi-2}
&\langle\varphi_p^\sigma|\sum_{\mu=1}^n B_\mu e_\mu|\varphi_q^\sigma\rangle \nonumber \\
&=-\delta_{pq}\sum_{\mu=1}^n B_\mu e_\mu +2\sum_{\mu=1}^n B_\mu \langle\phi_{p}^{\sigma\!,\mu}|\varphi_q^\sigma\rangle \nonumber \\
&+2\sum_{\mu=1}^n B_\mu e_\mu  \langle\phi_{p}^{\sigma\!,0}|\varphi_q^\sigma\rangle,
\end{align}
Inserting Eq.~\eqref{eq:orb_hc} gives:
\begin{align}
\label{eq:varphiXvarphi-2-2}
&2\sum_{\mu=1}^n B_\mu \langle\phi_{p}^{\sigma\!,\mu}|\varphi_q^\sigma\rangle \nonumber \\
&= 2\sum_{\mu=1}^n B_\mu \langle\phi_{p}^{\sigma\!,\mu}|\phi_q^{\sigma\!,0}\rangle +2\sum_{1\leq\mu,\nu\leq n} B_\nu \langle\phi_{p}^{\sigma\!,\nu}|\phi_q^{\sigma\!,\mu}\rangle e_\mu.
\end{align}
Similarly,
\begin{align}
\label{eq:varphiXvarphi-2-3}
&2\sum_{\mu=1}^n B_\mu e_\mu  \langle\phi_{p}^{\sigma\!,0}|\varphi_q^\sigma\rangle \nonumber \\
&= -2\sum_{\mu=1}^n B_\mu \langle\phi_{p}^{\sigma\!,0}|\phi_q^{\sigma\!, \mu} \rangle +2\sum_{\mu=1}^n B_\mu \langle\phi_{p}^{\sigma\!,0}|\phi_q^{\sigma\!, 0}\rangle e_\mu \nonumber \\
&+2 \sum_{1\leq\mu<\nu\leq n} \Big(B_\mu \langle\phi_{p}^{\sigma\!,0}|\phi_q^{\sigma\!, \nu}\rangle-B_\nu \langle\phi_{p}^{\sigma\!,0}|\phi_q^{\sigma\!, \mu}\rangle\Big)e_\mu e_\nu.
\end{align}
Due to Eq.~\eqref{eq:ortho2}, the first terms on the right-hand side of both Eqs.~\eqref{eq:varphiXvarphi-2-2} and \eqref{eq:varphiXvarphi-2-3} cancel each other, leading to:
\begin{align}
\label{eq:varphiXvarphi-2-4}
&\langle\varphi_p^\sigma|\sum_{\mu=1}^n B_\mu e_\mu|\varphi_q^\sigma\rangle \nonumber \\
&=\sum_{\mu=1}^n\Big[-\delta_{pq} B_\mu+2 B_\mu \langle\phi_{p}^{\sigma\!,0}|\phi_q^{\sigma\!, 0}\rangle +2\sum_{\nu=1}^n B_\nu \langle\phi_{p}^{\sigma\!,\nu}|\phi_q^{\sigma\!,\mu}\rangle \Big] e_\mu \nonumber \\
&+\sum_{1\leq\mu<\nu\leq n} 2\Big(B_\mu \langle\phi_{p}^{\sigma\!,0}|\phi_q^{\sigma\!, \nu}\rangle-B_\nu \langle\phi_{p}^{\sigma\!,0}|\phi_q^{\sigma\!, \mu}\rangle\Big)e_\mu e_\nu.
\end{align}
For the third term, first consider:
\begin{align}
\label{eq:varphiX-3}
&\overline{\varphi}_p^\sigma(\br)\sum_{1\leq\mu<\nu\leq n} C_{\mu\nu}e_\mu e_\nu \nonumber \\
&=\Big[\phi_{p}^{\sigma\!,0}(\br)-\sum_{\nu=1}^{n}\phi_{p}^{\sigma\!,\nu}(\br) e_\nu\Big]\sum_{1\leq\mu<\nu\leq n} C_{\mu\nu}e_\mu e_\nu \nonumber \\
& =\sum_{1\leq\mu<\nu\leq n} C_{\mu\nu}e_\mu e_\nu \overline{\varphi}_p^\sigma(\br)
+2\sum_{1\leq\mu<\nu\leq n} C_{\mu\nu} \phi_p^{\sigma\!,\mu}(\br)e_\nu \nonumber \\
&-2\sum_{1\leq\mu<\nu\leq n} C_{\mu\nu} \phi_p^{\sigma\!,\nu}(\br)e_\mu.
\end{align}
This leads to:
\begin{align}
\label{eq:varphiXvarphi-3}
&\langle\varphi_p^\sigma|\sum_{1\leq\mu<\nu\leq n} C_{\mu\nu}e_\mu e_\nu|\varphi_q^\sigma\rangle \nonumber \\
&=\delta_{pq}\sum_{1\leq\mu<\nu\leq n} C_{\mu\nu}e_\mu e_\nu \nonumber \\ &+2\sum_{1\leq\mu<\nu\leq n} C_{\mu\nu} e_\nu \langle\phi_{p}^{\sigma\!,\mu}|\varphi_q^\sigma\rangle \nonumber \\
&-2\sum_{1\leq\mu<\nu\leq n} C_{\mu\nu}e_\mu \langle\phi_{p}^{\sigma\!,\nu}|\varphi_q^\sigma\rangle,
\end{align}
Inserting Eq.~\eqref{eq:orb_hc} gives:
\begin{align}
\label{eq:varphiXvarphi-3-2}
&2\sum_{1\leq\mu<\nu\leq n} C_{\mu\nu} e_\nu \langle\phi_{p}^{\sigma\!,\mu}|\varphi_q^\sigma\rangle \nonumber \\
&=-2\sum_{1\leq\mu<\nu\leq n} C_{\mu\nu} \langle\phi_{p}^{\sigma\!,\mu}|\phi_q^{\sigma\!,\nu}\rangle \nonumber \\
&+2\sum_{1\leq\mu<\nu\leq n} C_{\mu\nu} \langle\phi_{p}^{\sigma\!,\mu}|\phi_q^{\sigma\!,0}\rangle e_\nu \nonumber \\
&-2\sum_{1\leq\mu<\nu\leq n} C_{\mu\nu} \langle\phi_{p}^{\sigma\!,\mu}|\phi_q^{\sigma\!,\mu}\rangle e_\mu e_\nu \nonumber \\
&-2\sum_{1\leq\lambda<\mu<\nu\leq n} C_{\mu\nu} \langle\phi_{p}^{\sigma\!,\mu}|\phi_q^{\sigma\!,\lambda}\rangle e_\lambda e_\nu \nonumber \\
&+2\sum_{1\leq\lambda<\mu<\nu\leq n} \Big( C_{\lambda\mu} \langle\phi_{p}^{\sigma\!,\lambda}|\phi_q^{\sigma\!,\nu}\rangle-C_{\lambda\nu} \langle\phi_{p}^{\sigma\!,\lambda}|\phi_q^{\sigma\!,\mu}\rangle\Big) e_\mu e_\nu
\end{align}
Similarly,
\begin{align}
\label{eq:varphiXvarphi-3-3}
&-2\sum_{1\leq\mu<\nu\leq n} C_{\mu\nu} e_\mu \langle\phi_{p}^{\sigma\!,\nu}|\varphi_q^\sigma\rangle \nonumber \\
&=2\sum_{1\leq\mu<\nu\leq n} C_{\mu\nu} \langle\phi_{p}^{\sigma\!,\nu}|\phi_q^{\sigma\!,\mu}\rangle \nonumber \\
&-2\sum_{1\leq\mu<\nu\leq n} C_{\mu\nu} \langle\phi_{p}^{\sigma\!,\nu}|\phi_q^{\sigma\!,0}\rangle e_\mu \nonumber \\
&-2\sum_{1\leq\mu<\nu\leq n} C_{\mu\nu} \langle\phi_{p}^{\sigma\!,\nu}|\phi_q^{\sigma\!,\nu}\rangle e_\mu e_\nu \nonumber \\
&-2\sum_{1\leq\lambda<\mu<\nu\leq n} C_{\lambda \mu} \langle\phi_{p}^{\sigma\!,\mu}|\phi_q^{\sigma\!,\nu}\rangle e_\lambda e_\nu \nonumber \\
&+2\sum_{1\leq\lambda<\mu<\nu\leq n} \Big( C_{\mu\nu} \langle\phi_{p}^{\sigma\!,\nu}|\phi_q^{\sigma\!,\lambda}\rangle-C_{\lambda\nu} \langle\phi_{p}^{\sigma\!,\nu}|\phi_q^{\sigma\!,\mu}\rangle\Big) e_\lambda e_\mu
\end{align}
By substituting Eqs.~\eqref{eq:varphiXvarphi-3-2} and \eqref{eq:varphiXvarphi-3-3} into Eq.~\eqref{eq:varphiXvarphi-3}, the expression becomes:
\begin{align}
\label{eq:varphiXvarphi-3-4}
&\langle\varphi_p^\sigma|\sum_{1\leq\mu<\nu\leq n} C_{\mu\nu}e_\mu e_\nu|\varphi_q^\sigma\rangle \nonumber \\
&=\delta_{pq}\sum_{1\leq\mu<\nu\leq n} C_{\mu\nu}e_\mu e_\nu \nonumber \\
&+\sum_{\mu=1}^n \Big(-2 \sum_{\nu=1}^n C_{\mu\nu} \langle\phi_{p}^{\sigma\!,\nu}|\phi_q^{\sigma\!,0}\rangle\Big) e_\mu \nonumber \\
&+\sum_{1\leq\mu<\nu\leq n}2\sum_{\lambda=1}^n\Big( C_{\lambda\mu} \langle\phi_{p}^{\sigma\!,\lambda}|\phi_q^{\sigma\!,\nu}\rangle-C_{\lambda\nu} \langle\phi_{p}^{\sigma\!,\lambda}|\phi_q^{\sigma\!,\mu}\rangle\Big) e_\mu e_\nu.
\end{align}
Reaching this result requires several operations on Eqs.~\eqref{eq:varphiXvarphi-3-2} and \eqref{eq:varphiXvarphi-3-3}, utilizing Eq.~\eqref{eq:ortho2} and the antisymmetry of $C$. Specifically, the first terms in Eqs.~\eqref{eq:varphiXvarphi-3-2} and \eqref{eq:varphiXvarphi-3-3} cancel each other out. The second terms combine to form the second term in Eq.~\eqref{eq:varphiXvarphi-3-4}, and the remaining terms from Eqs.~\eqref{eq:varphiXvarphi-3-2} and \eqref{eq:varphiXvarphi-3-3} contribute to the third term in Eq.~\eqref{eq:varphiXvarphi-3-4}.

Next, by inserting Eqs.~\eqref{eq:varphiXvarphi-1}, \eqref{eq:varphiXvarphi-2-4}, and \eqref{eq:varphiXvarphi-3-4} into Eq.~\eqref{eq:varphiXvarphi}, Eq.~\eqref{eq:Xp} is obtained, with:
\begin{equation}
\label{eq:Ap}
A'=\delta_{pq}A,
\end{equation}
\begin{align}
\label{eq:Bp}
B'_\mu&=-\delta_{pq} B_\mu+2 B_\mu \langle\phi_{p}^{\sigma\!,0}|\phi_q^{\sigma\!, 0}\rangle \nonumber \\
&+2\sum_{\nu=1}^n B_\nu \langle\phi_{p}^{\sigma\!,\nu}|\phi_q^{\sigma\!,\mu}\rangle 
-2 \sum_{\nu=1}^n C_{\mu\nu} \langle\phi_{p}^{\sigma\!,\nu}|\phi_q^{\sigma\!,0}\rangle,
\end{align}
and
\begin{align}
\label{eq:Cp}
C'_{\mu\nu} &=\delta_{pq} C_{\mu\nu}+2 \Big(B_\mu \langle\phi_{p}^{\sigma\!,0}|\phi_q^{\sigma\!, \nu}\rangle-B_\nu \langle\phi_{p}^{\sigma\!,0}|\phi_q^{\sigma\!, \mu}\rangle\Big) \nonumber \\
&+2\sum_{\lambda=1}^n\Big( C_{\lambda\mu} \langle\phi_{p}^{\sigma\!,\lambda}|\phi_q^{\sigma\!,\nu}\rangle-C_{\lambda\nu} \langle\phi_{p}^{\sigma\!,\lambda}|\phi_q^{\sigma\!,\mu}\rangle\Big),
\end{align}
Thus, it follows that $C'_{\mu\nu} = -C'_{\nu\mu}$. \(\square\)

Lemma 1 is interesting because any quadratic quantity in $\{e_1, e_2, \cdots, e_n\}$ can be written in the form of Eq.~\eqref{eq:X}. By repeatedly applying the bra-ket integrals over different orbitals, this ultimately results in a quadratic form as well. As such, the multiple integrals in Eq.~\eqref{eq:spatialint2} also yield a quadratic form, i.e.,
\begin{align}
\label{eq:spatialint5}
&\mathcal{O}_{(i_1i_2\cdots i_N),k}^{(i'_1i'_2\cdots i'_N)}\!(\br) \nonumber \\
&=\Big[\mathcal{A}_{(i_k i_{k+1}\cdots i_N)}^{(i'_k i'_{k+1}\cdots i'_N)}\!(\br)+\sum_{\mu=1}^n \mathcal{B}_{(i_k i_{k+1}\cdots i_N),\mu}^{(i'_k i'_{k+1}\cdots i'_N)}\!(\br) e_\mu \nonumber \\
&+\sum_{1\leq\mu<\nu\leq n} \mathcal{C}_{(i_k i_{k+1}\cdots i_N),\mu\nu}^{(i'_k i'_{k+1}\cdots i'_N)}\!(\br) e_\mu e_\nu\Big]\prod_{j=1}^{k-1}\delta_{i_ji'_j},
\end{align}
where 
\begin{equation}
\label{eq:spatialint6}
\mathcal{A}_{(i_k i_{k+1}\cdots i_N)}^{(i'_k i'_{k+1}\cdots i'_N)}\!(\br)=\mathcal{A}_{i_k}^{i'_k}\!(\br)\prod_{j=k+1}^{N}\delta_{i_ji'_j}.
\end{equation}
Therefore, when $(i_1 i_2 \cdots i_N) = (i'_1 i'_2 \cdots i'_N)$, $\mathcal{O}_{(i_1 i_2 \cdots i_N),k}^{(i'_1 i'_2 \cdots i'_N)}\!(\br)$ reduces to Eq.~\eqref{eq:spatialint4}. When $(i_1 i_2 \cdots i_N) \neq (i'_1 i'_2 \cdots i'_N)$, there are at least two indices differing between them, so the scalar term disappears in $\mathcal{O}_{(i_1 i_2 \cdots i_N),k}^{(i'_1 i'_2 \cdots i'_N)}\!(\br)$, leading to
\begin{align}
\label{eq:spatialint7}
&\mathcal{O}_{(i_1i_2\cdots i_N),k}^{(i'_1i'_2\cdots i'_N)}\!(\br) \nonumber \\
&=\Big[\sum_{\mu=1}^n \mathcal{B}_{(i_k i_{k+1}\cdots i_N),\mu}^{(i'_k i'_{k+1}\cdots i'_N)}\!(\br) e_\mu \nonumber \\
&+\sum_{1\leq\mu<\nu\leq n} \mathcal{C}_{(i_k i_{k+1}\cdots i_N),\mu\nu}^{(i'_k i'_{k+1}\cdots i'_N)}\!(\br) e_\mu e_\nu\Big]\prod_{j=1}^{k-1}\delta_{i_ji'_j}.
\end{align}
Because the summations in Eq.~\eqref{eq:phirhophi_2} run over all permutations of $(i_1 i_2 \cdots i_N)$ and $(i'_1 i'_2 \cdots i'_N)$, for each $\mathcal{O}_{(i_1 i_2 \cdots i_N),k}^{(i'_1 i'_2 \cdots i'_N)}\!(\br)$, there exists $\mathcal{O}_{(i'_1 i'_2 \cdots i'_N),k}^{(i_1 i_2 \cdots i_N)}\!(\br)$. And, it holds that
\begin{align}
\label{eq:overlineO}
&\mathcal{O}_{(i'_1i'_2\cdots i'_N),k}^{(i_1i_2\cdots i_N)}\!(\br)=\overline{\mathcal{O}}_{(i_1i_2\cdots i_N),k}^{(i'_1i'_2\cdots i'_N)}\!(\br) \nonumber \\
&=\Big[-\sum_{\mu=1}^n \mathcal{B}_{(i_k i_{k+1}\cdots i_N),\mu}^{(i'_k i'_{k+1}\cdots i'_N)}\!(\br) e_\mu \nonumber \\
&-\sum_{1\leq\mu<\nu\leq n} \mathcal{C}_{(i_k i_{k+1}\cdots i_N),\mu\nu}^{(i'_k i'_{k+1}\cdots i'_N)}\!(\br) e_\mu e_\nu\Big]\prod_{j=1}^{k-1}\delta_{i_ji'_j}.
\end{align}
Thus, $\mathcal{O}_{(i_1 i_2 \cdots i_N),k}^{(i'_1 i'_2 \cdots i'_N)}\!(\br)$ and $\mathcal{O}_{(i'_1 i'_2 \cdots i'_N),k}^{(i_1 i_2 \cdots i_N)}\!(\br)$ cancel out when $(i_1 i_2 \cdots i_N) \neq (i'_1 i'_2 \cdots i'_N)$. Only the terms for which $(i_1 i_2 \cdots i_N) = (i'_1 i'_2 \cdots i'_N)$ survive. By substituting Eq.~\eqref{eq:spatialint4} into Eq.~\eqref{eq:phirhophi_2}, the density and spin density expressions in Eqs.~\eqref{eq:density} and \eqref{eq:spinden} are obtained. Following the same procedure, the kinetic energy in Eq.~\eqref{eq:Tn} and the external energy in Eq.~\eqref{eq:totalene} can be derived.

Note that there are other ways to extend the orbitals to hypercomplex number systems beyond Eq.~\eqref{eq:orb_hc}, such as including quadratic or higher-order terms in $\{e_1, e_2, \cdots, e_n\}$ \cite{HC1989,CA2019}. However, the uniqueness of Eq.~\eqref{eq:orb_hc} lies in the fact that Lemma 1 ensures the density, kinetic energy, and external energy are real, and they can be conveniently formulated in the forms of Eqs.~\eqref{eq:density}, \eqref{eq:Tn}, and \eqref{eq:totalene}. In contrast, higher-order terms of $\{e_\mu\}$ in the spatial orbitals do not provide such a guarantee.

\section{\label{sec4} Extension of Hypercomplex Orbital Representation}

To further enrich HCOFT, this section introduces an alternative definition for hypercomplex orbitals. Rather than incorporating higher-order terms in $\{e_1, e_2, \cdots, e_n\}$, the basis is extended by introducing additional types of basis elements within the framework of Clifford algebra \cite{Lounesto2001,Selig2005}. Specifically, in addition to the original set $\{e_1, e_2, \cdots, e_n\}$, which satisfies the relation in Eq.~\eqref{eq:CA}, two supplementary sets of basis elements are introduced: the set $\{e_{n+1}, e_{n+2}, \cdots, e_{n+m}\}$, where each element squares to 1, and the set $\{e_{n+m+1}, e_{n+m+2}, \cdots, e_{n+m+l}\}$, where each element squares to 0. These basis elements satisfy the following relations \cite{Lounesto2001,Selig2005}:
\begin{align}
\label{eq:CA2}
e_\mu^2=\begin{cases}
-1, & 1\leq \mu\leq n, \\
1,  & n+1\leq \mu\leq n+m, \\
0,  & n+m+1\leq\mu\leq n+m+l,
\end{cases}
\end{align}
as well as the anti-commutation relation:
\begin{equation}
\label{eq:CA3}
e_\mu e_\nu=-e_\nu e_\mu,  1\leq \mu\neq\nu\leq n+m+l.
\end{equation}
Using this extended basis, the hypercomplex orbitals $\{\varphi_p^\sigma\}$ are defined as
\begin{equation}
\label{eq:orb_hc2}
\varphi_p^\sigma(\br)=\phi_{p}^{\sigma\!,0}(\br)+\sum_{\mu=1}^{n+m+l}\phi_{p}^{\sigma\!,\mu}(\br) e_\mu,
\end{equation}
while their Clifford conjugates are given by
\begin{equation}
\label{eq:orb_conj2}
\overline{\varphi}_p^\sigma(\br)=\phi_{p}^{\sigma\!,0}(\br)-\sum_{\mu=1}^{n+m+l}\phi_{p}^{\sigma\!,\mu}(\br) e_\mu,
\end{equation}
Here, the set $\{\phi_{p}^{\sigma,\mu}\}$ consists of real functions as before. 

By inserting Eqs.~\eqref{eq:orb_hc2} and \eqref{eq:orb_conj2} into Eqs.~\eqref{eq:ortho} and \eqref{eq:iden}, the following orthogonality condition is obtained:
\begin{align}
\label{eq:ortho1-2}
& \sum_{\mu=0}^n \langle\phi_p^{\sigma,\mu}|\phi_q^{\sigma,\mu}\rangle-\sum_{\mu=n+1}^{n+m} \langle\phi_p^{\sigma,\mu}|\phi_q^{\sigma,\mu}\rangle\nonumber \\
&+\sum_{\mu=1}^{n+m+l} (\langle\phi_p^{\sigma,0}|\phi_q^{\sigma,\mu}\rangle-\langle\phi_p^{\sigma,\mu}|\phi_q^{\sigma,0}\rangle)e_\mu \nonumber \\
&+\sum_{1\leq\mu<\nu\leq n+m+l} (\langle\phi_p^{\sigma,\nu}|\phi_q^{\sigma,\mu}\rangle-\langle\phi_p^{\sigma,\mu}|\phi_q^{\sigma,\nu}\rangle)e_\mu e_\nu=\delta_{pq}.
\end{align}
and the completeness relation takes the form:
\begin{align}
\label{eq:iden1-2}
& \sum_{\mu=0}^n\sum_p |\phi_p^{\sigma,\mu}\rangle\langle\phi_p^{\sigma,\mu}|- \sum_{\mu=n+1}^{n+m}\sum_p |\phi_p^{\sigma,\mu}\rangle\langle\phi_p^{\sigma,\mu}| \nonumber \\
&+\sum_{\mu=1}^{n+m+l}\sum_p (|\phi_p^{\sigma,\mu}\rangle\langle\phi_p^{\sigma,0}|-|\phi_p^{\sigma,0}\rangle\langle\phi_p^{\sigma,\mu}|)e_\mu \nonumber \\
&+\sum_{1\leq\mu<\nu\leq n+m+l}\sum_p (|\phi_p^{\sigma,\mu}\rangle\langle\phi_p^{\sigma,\nu}|-|\phi_p^{\sigma,\nu}\rangle\langle\phi_p^{\sigma,\mu}|)e_\mu e_\nu \nonumber \\
&=\hat{I}_K.
\end{align}
These lead to the following conditions:  
For orthogonality,
\begin{align}
\label{eq:ortho2-2}
\begin{cases}
\sum_{\mu=0}^n \langle\phi_p^{\sigma,\mu}|\phi_q^{\sigma,\mu}\rangle-\sum_{\mu=n+1}^{n+m} \langle\phi_p^{\sigma,\mu}|\phi_q^{\sigma,\mu}\rangle=\delta_{pq},  \\
\langle\phi_p^{\sigma,\mu}|\phi_q^{\sigma,\nu}\rangle=\langle\phi_p^{\sigma,\nu}|\phi_q^{\sigma,\mu}\rangle,  
\end{cases}
\end{align}
For the completeness relation,
\begin{align}
\label{eq:iden2-2}
\begin{cases}
\sum_{\mu=0}^n \sum_p |\phi_p^{\sigma,\mu}\rangle\langle\phi_p^{\sigma,\mu}|-\sum_{\mu=n+1}^{n+m} \sum_p |\phi_p^{\sigma,\mu}\rangle\langle\phi_p^{\sigma,\mu}|=\hat{I}_K,  \\
\sum_p|\phi_p^{\sigma,\mu}\rangle\langle\phi_p^{\sigma,\nu}|=\sum_p|\phi_p^{\sigma,\nu}\rangle\langle\phi_p^{\sigma,\mu}|,  
\end{cases}
\end{align}

The set \(\{\phi_{p}^{\sigma,\mu}\}\) can be expanded in terms of a set of orthonormal functions \(\{\xi_p\}\), similar to Eq.~\eqref{eq:phi}, as follows:
\begin{equation}
\label{eq:phi-2}
\phi_p^{\sigma,\mu}(\textbf{r}) = \sum_q \xi_q(\textbf{r}) V_{pq}^{\sigma,\mu},
\end{equation}
where \(\{{V}^{\sigma,\mu}\}\) represents a set of \(1\!+\!n\!+\!m\!+l\) \(K \times K\) matrices, with each \({V}^{\sigma,\mu}\) containing the expansion coefficients for the \(\mu\)-th component of the hypercomplex orbitals.  
Using this representation, Eqs.~\eqref{eq:ortho2-2} and \eqref{eq:iden2-2} can be rewritten in terms of \(\{{V}^{\sigma,\mu}\}\) as follows:  For the orthogonality condition,
\begin{align}
\label{eq:cond_hc1-2}
\begin{cases}
\sum_{\mu=0}^{n} {V}^{\sigma,\mu}{V}^{\sigma,\mu T}-\sum_{\mu=n+1}^{n+m} {V}^{\sigma,\mu}{V}^{\sigma,\mu T} = \hat{I}_K,  \\
{V}^{\sigma,\mu}{V}^{\sigma,\nu T} = {V}^{\sigma,\nu}{V}^{\sigma,\mu T}.
\end{cases}
\end{align}
For the completeness relation,  
\begin{align}
\label{eq:cond_hc2-2}
\begin{cases}
\sum_{\mu=0}^{n} {V}^{\sigma,\mu T}{V}^{\sigma,\mu}-\sum_{\mu=n+1}^{n+m} {V}^{\sigma,\mu T}{V}^{\sigma,\mu} = {I}_K,  \\
{V}^{\sigma,\mu T}{V}^{\sigma,\nu} = {V}^{\sigma,\nu T}{V}^{\sigma,\mu}.
\end{cases}
\end{align}

Next, it is necessary to verify whether the newly introduced definition of hypercomplex orbitals, i.e., Eq.~\eqref{eq:orb_hc2}, along with the resulting wavefunction \(|\Phi\rangle\) from Eq.~\eqref{eq:dete}, still leads to the same expressions for the density in Eq.~\eqref{eq:density}, the spin density in Eq.~\eqref{eq:spinden}, the kinetic energy in Eq.~\eqref{eq:Tn}, and the external energy in Eq.~\eqref{eq:totalene}. 

Since the Clifford conjugate reverses the order in a product of hypercomplex orbitals \cite{Lounesto2001}, as indicated by Eq.~\eqref{eq:prod_conj}, the expression for \(\langle\Phi|\) retains the same form as Eq.~\eqref{eq:dete_l}. Furthermore, the density can still be formulated in the same structure as Eq.~\eqref{eq:phirhophi_2}, with the spin part remaining unchanged as in Eq.~\eqref{eq:spinint}, and the spatial part maintaining the same form as in Eq.~\eqref{eq:spatialint2}. These properties hold because, up to this point, the modifications in the definition of hypercomplex orbitals have not altered the fundamental structure of these equations.

Substituting the new definition of hypercomplex orbitals, Eq.~\eqref{eq:orb_hc2}, Eq.~\eqref{eq:int_k} now becomes
\begin{align}
\label{eq:int_k2}
&\int\!\overline{\varphi}_{\tilde{i}_k}^{\sigma_{i_k}}\!(\!\br_k\!)\!\delta(\br-\br_k)\!\varphi_{\tilde{i}'_{k}}^{\sigma_{i'_k}}\!(\!\br_k\!)d\br_k\nonumber \\
&=\overline{\varphi}_{\tilde{i}_k}^{\sigma_{i_k}}\!(\!\br\!)\!\varphi_{\tilde{i}'_{k}}^{\sigma_{i'_k}}\!(\!\br\!) \nonumber \\
&=\mathcal{A}_{i_k}^{i'_k}\!(\br)\!+\!\sum_{\mu=1}^{n+m+l}\!\mathcal{B}_{i_k,\mu}^{i'_k}\!(\br)e_\mu \nonumber \\
&+ \sum_{1\leq\mu<\nu\leq n+m+l}\!\mathcal{C}_{i_k,\mu\nu}^{i'_k}\!(\br)e_\mu e_\nu,
\end{align}
where 
\begin{equation}
\label{eq:Ak-2}
\mathcal{A}_{i_k}^{i'_k}\!(\br)=\sum_{\mu=0}^n \phi_{\tilde{i}_k}^{\sigma_{i_k},\mu}\!(\br)\phi_{\tilde{i}'_k}^{\sigma_{i'_k},\mu}\!(\br)-\sum_{\mu=n+1}^{n+m} \phi_{\tilde{i}_k}^{\sigma_{i_k},\mu}\!(\br)\phi_{\tilde{i}'_k}^{\sigma_{i'_k},\mu}\!(\br),
\end{equation}
\begin{equation}
\label{eq:Bk-2}
\mathcal{B}_{i_k,\mu}^{i'_k}\!(\br)= \phi_{\tilde{i}_k}^{\sigma_{i_k},0}\!(\br)\phi_{\tilde{i}'_k}^{\sigma_{i'_k},\mu}\!(\br)- \phi_{\tilde{i}_k}^{\sigma_{i_k},\mu}\!(\br)\phi_{\tilde{i}'_k}^{\sigma_{i'_k},0}\!(\br),
\end{equation}
and
\begin{equation}
\label{eq:Ck-2}
\mathcal{C}_{i_k,\mu\nu}^{i'_k}\!(\br)= \phi_{\tilde{i}_k}^{\sigma_{i_k},\nu}\!(\br)\phi_{\tilde{i}'_k}^{\sigma_{i'_k},\mu}\!(\br)- \phi_{\tilde{i}_k}^{\sigma_{i_k},\mu}\!(\br)\phi_{\tilde{i}'_k}^{\sigma_{i'_k},\nu}\!(\br).
\end{equation}
When \(i_k = i'_k\), both \(\mathcal{B}_{i_k,\mu}^{i_k}\!(\br)\) and \(\mathcal{C}_{i_k,\mu\nu}^{i_k}\!(\br)\) vanish, reducing Eq.~\eqref{eq:int_k2} to a real function that commutes with any hypercomplex orbital. Consequently, this results in:
\begin{equation}
\label{eq:spatialint3-2}
\mathcal{O}_{(i_1i_2\cdots i_N),k}^{(i'_1i'_2\cdots i'_N)}\!(\br) =\mathcal{O}_{(i_1i_2\cdots i_N),k}^{(i_1i_2\cdots i_N)}\!(\br)\prod_{j=1}^{N}\delta_{i_ji'_j},
\end{equation}
which implies that only the terms with \((i_1i_2\cdots i_N) = (i'_1i'_2\cdots i'_N)\) contribute in this case. Additionally, 
\begin{equation}
\label{eq:spatialint4-2}
\mathcal{O}_{(i_1i_2\cdots i_N),k}^{(i_1i_2\cdots i_N)}\!(\br)=\overline{\varphi}_{\tilde{i}_k}^{\sigma_{i_k}}(\br)\varphi_{\tilde{i}_k}^{\sigma_{i_k}}(\br)
=\mathcal{A}_{i_k}^{i_k}\!(\br).
\end{equation}

Similar to Lemma 1, the new set of hypercomplex orbitals satisfies the following lemma:

\textbf{Lemma 2:} Let $X$ be hypercomplex, of the form:
\begin{equation}
\label{eq:X-2}
X=A+\sum_{\mu=1}^{n+m+l} B_\mu e_\mu +\sum_{1\leq\mu<\nu\leq n+m+l} C_{\mu\nu}e_\mu e_\nu,
\end{equation}
where $A$, $B_\mu$, and $C_{\mu\nu}$ are real valued, and $C_{\mu\nu} = -C_{\nu\mu}$. Then, the bra-ket integral over two hypercomplex spatial orbitals, that is $\langle \varphi_p^\sigma | X | \varphi_q^\sigma \rangle$, retains the same hypercomplex structure:
\begin{equation}
\label{eq:Xp-2}
\langle\varphi_p^\sigma|X|\varphi_q^\sigma\rangle=A'+\sum_{\mu=1}^{n+m+l} B'_\mu e_\mu +\sum_{1\leq\mu<\nu\leq n+m+l} C'_{\mu\nu}e_\mu e_\nu,
\end{equation}
where $A'$, $B'_\mu$, and $C'_{\mu\nu}$ are real valued satisfying the antisymmetry condition $C'_{\mu\nu} = -C'_{\nu\mu}$. Moreover, the scalar component transforms as:
\begin{equation}
A' = \delta_{pq} A.
\end{equation}

\textbf{Proof 2:} By inserting Eq.~\eqref{eq:X-2} into Eq.~\eqref{eq:Xp-2}, the following expression is obtained:
\begin{align}
\label{eq:varphiXvarphi2}
\langle\varphi_p^\sigma|X|\varphi_q^\sigma\rangle&=\langle\varphi_p^\sigma|A|\varphi_q^\sigma\rangle \nonumber \\
&+\langle\varphi_p^\sigma|\sum_{\mu=1}^n B_\mu e_\mu|\varphi_q^\sigma\rangle \nonumber \\
&+\langle\varphi_p^\sigma|\sum_{1\leq\mu<\nu\leq n} C_{\mu\nu}e_\mu e_\nu|\varphi_q^\sigma\rangle,
\end{align}
Following the derivations in Eqs.~\eqref{eq:varphiXvarphi-1}--\eqref{eq:varphiXvarphi-3-4}, the three terms on the right-hand side of Eq.~\eqref{eq:varphiXvarphi2} yield the results:
\begin{equation}
\label{eq:varphiXvarphi-1-2}
\langle\varphi_p^\sigma|A|\varphi_q^\sigma\rangle=\delta_{pq}A,
\end{equation}
\begin{align}
\label{eq:varphiXvarphi-2-4-2}
&\langle\varphi_p^\sigma|\sum_{\mu=1}^{n+m+l} B_\mu e_\mu|\varphi_q^\sigma\rangle \nonumber \\
&=\sum_{\mu=1}^{n+m+l}\Big[-\delta_{pq} B_\mu+2 B_\mu \langle\phi_{p}^{\sigma\!,0}|\phi_q^{\sigma\!, 0}\rangle \nonumber \\
&\ \ \ \ \ \ \ \ \ \ \ \ \ \ -2\sum_{\nu=1}^{n+m+l} e_\nu^2 B_\nu \langle\phi_{p}^{\sigma\!,\nu}|\phi_q^{\sigma\!,\mu}\rangle \Big] e_\mu \nonumber \\
&+\sum_{1\leq\mu<\nu\leq n\!+\!m\!+\!l} 2\Big(B_\mu \langle\phi_{p}^{\sigma\!,0}|\phi_q^{\sigma\!, \nu}\rangle-B_\nu \langle\phi_{p}^{\sigma\!,0}|\phi_q^{\sigma\!, \mu}\rangle\Big)e_\mu e_\nu,
\end{align}
and 
\begin{align}
\label{eq:varphiXvarphi-3-4-2}
&\langle\varphi_p^\sigma|\sum_{1\leq\mu<\nu\leq n+m+l} C_{\mu\nu}e_\mu e_\nu|\varphi_q^\sigma\rangle \nonumber \\
&=\delta_{pq}\sum_{1\leq\mu<\nu\leq n+m+l} C_{\mu\nu}e_\mu e_\nu \nonumber \\
&+\sum_{\mu=1}^{n+m+l} \Big(2 \sum_{\nu=1}^{n+m+l} e_\nu^2 C_{\mu\nu} \langle\phi_{p}^{\sigma\!,\nu}|\phi_q^{\sigma\!,0}\rangle\Big) e_\mu \nonumber \\
&-\sum_{1\leq\mu<\nu\leq {n+m+l}}2\sum_{\lambda=1}^{n+m+l}e_\lambda^2\Big( C_{\lambda\mu} \langle\phi_{p}^{\sigma\!,\lambda}|\phi_q^{\sigma\!,\nu}\rangle \nonumber \\
&\ \ \ \ \ \ \ \ \ \ \ \ \ \ \ \ \ \ \ \ \ \ \ \ \ \ \ \ \ \ \ \ \ \ -C_{\lambda\nu} \langle\phi_{p}^{\sigma\!,\lambda}|\phi_q^{\sigma\!,\mu}\rangle\Big) e_\mu e_\nu.
\end{align}
By substituting Eqs.~\eqref{eq:varphiXvarphi-1-2}, \eqref{eq:varphiXvarphi-2-4-2}, and \eqref{eq:varphiXvarphi-3-4-2} into Eq.~\eqref{eq:varphiXvarphi2}, Eq.~\eqref{eq:Xp-2} is obtained, with 
\begin{equation}
\label{eq:Ap-2}
A'=\delta_{pq}A,
\end{equation}
\begin{align}
\label{eq:Bp-2}
&B'_\mu=-\delta_{pq} B_\mu+2 B_\mu \langle\phi_{p}^{\sigma\!,0}|\phi_q^{\sigma\!, 0}\rangle \nonumber \\
&-2\sum_{\nu=1}^n e_\nu^2 B_\nu \langle\phi_{p}^{\sigma\!,\nu}|\phi_q^{\sigma\!,\mu}\rangle 
+2 \sum_{\nu=1}^n e_\nu^2 C_{\mu\nu} \langle\phi_{p}^{\sigma\!,\nu}|\phi_q^{\sigma\!,0}\rangle,
\end{align}
and
\begin{align}
\label{eq:Cp-2}
&C'_{\mu\nu} =\delta_{pq} C_{\mu\nu}+2 \Big(B_\mu \langle\phi_{p}^{\sigma\!,0}|\phi_q^{\sigma\!, \nu}\rangle-B_\nu \langle\phi_{p}^{\sigma\!,0}|\phi_q^{\sigma\!, \mu}\rangle\Big) \nonumber \\
&-2\sum_{\lambda=1}^{n+m+l} e_\lambda^2 \Big( C_{\lambda\mu} \langle\phi_{p}^{\sigma\!,\lambda}|\phi_q^{\sigma\!,\nu}\rangle-C_{\lambda\nu} \langle\phi_{p}^{\sigma\!,\lambda}|\phi_q^{\sigma\!,\mu}\rangle\Big),
\end{align}
which ensures that \(C'_{\mu\nu}=-C'_{\nu\mu}\). \(\square\)

Then, following the same discussions from the previous section, the density can be formulated in the form of Eqs.~\eqref{eq:density} and \eqref{eq:spinden}. Furthermore, by inserting Eqs.~\eqref{eq:orb_hc2}, \eqref{eq:orb_conj2}, and \eqref{eq:phi-2} into Eq.~\eqref{eq:spinden}, the following expression is obtained:
\begin{align}
\label{eq:rho_hc-2}
&\rho^{\sigma}_{\{\chi_p^\sigma, \lambda_p^\sigma\}}(\br) =\sum_{i=1}^{N_\sigma}\overline{\varphi}_{i}^{\sigma}(\br)\varphi_{i}^{\sigma}(\br) \nonumber \\
&= \sum_i^{N_\sigma} \Big(\sum_{\mu=0}^n |\phi_{i}^{\sigma,\mu}(\mathbf{r})|^2-\sum_{\mu=n+1}^{n+m} |\phi_{i}^{\sigma,\mu}(\mathbf{r})|^2 \Big) \nonumber \\
&= \sum_{p=1}^{K}{\lambda}_{p}^{\sigma} |\chi_p^{\sigma}(\br)|^2.
\end{align}
Here, HCOs and their occupations \( \{\chi_p^\sigma, \lambda_p^\sigma\} \) correspond to the eigenvectors and eigenvalues of \( D^\sigma \),
\begin{equation}
\label{eq:D_hc-2}
D^\sigma = \sum_{\mu=0}^{n} {V}^{\sigma,\mu T} {\rm{I}}^{N_\sigma}_K {V}^{\sigma,\mu}-\sum_{\mu=n+1}^{n+m} {V}^{\sigma,\mu T} {\rm{I}}^{N_\sigma}_K {V}^{\sigma,\mu}.
\end{equation}
Following the same procedure, the kinetic energy from Eq.~\eqref{eq:Tn} can be obtained as:
\begin{align}
\label{eq:Tn-2}
&T_n[\{\chi_p^\sigma, \lambda_p^\sigma\}]=-\frac{1}{2}\sum_\sigma\sum_{i=1}^{N_\sigma}\langle\varphi_{i}^{\sigma}| \nabla^2|\varphi_{i}^{\sigma}\rangle \nonumber \\
&=-\frac{1}{2}\sum_\sigma\sum_{i=1}^{N_\sigma}\Big(\sum_{\mu=0}^n \langle\phi_{i}^{\sigma,\mu}| \nabla^2|\phi_{i}^{\sigma,\mu}\rangle-\sum_{\mu=n+1}^{n+m} \langle\phi_{i}^{\sigma,\mu}| \nabla^2|\phi_{i}^{\sigma,\mu}\rangle\Big) \nonumber \\
&=-\frac{1}{2}\sum_{\sigma}\sum_{p} \lambda_{p}^{\sigma} \langle \chi_p^\sigma | \nabla^2 | \chi_p^{\sigma} \rangle.
\end{align}

With these derivations completed, it is intriguing to compare the results obtained from different definitions of hypercomplex orbitals. Although the extended basis in Eq.~\eqref{eq:orb_hc2} retains several beneficial properties of the previously defined hypercomplex orbitals in Eq.~\eqref{eq:orb_hc}---particularly, Lemma 2 ensures that the density and kinetic energy remain real and can be conveniently expressed in the forms of Eqs.~\eqref{eq:rho_hc-2} and \eqref{eq:Tn-2}---it does not guarantee the positive definiteness of \( D^\sigma \) in Eq.~\eqref{eq:D_hc-2}. As a result, this formulation does not necessarily constrain the occupations of HCOs within the range \([0,1]\). This characteristic may suggest potential applications in the study of alternative physical systems, such as those involving antiparticles or electronic holes, though further theoretical validation is required. Moreover, this observation highlights the distinctive nature of Eq.~\eqref{eq:orb_hc}, which not only preserves the desired physical properties but also enables the use of a single-reference framework for describing multi-reference systems.

\section{\label{sec:int} Conclusions}

This work provides a formal derivation of the functional theory with hypercomplex orbitals, aiming to enhance the understanding of key concepts and foster the development of novel methods. Recent study has demonstrated that hypercomplex orbitals with integer occupations are equivalent to real-valued hierarchically correlated orbitals (HCOs) that allow fractional occupations, when considered as fundamental descriptors of many-electron systems, from which the HCO functional theory (HCOFT) is derived.  Building on prior research, this work offers a detailed mathematical exposition of HCOFT, outlining its foundational principles and functional formulation. To address the complexity introduced by the hypercomplex representation in the derivation of fundamental quantities, the algebraic properties of Clifford algebra are carefully applied to rigorously prove two key lemmas for calculating the integrals of hypercomplex orbitals. Leveraging these results, it is shown that, despite the added complexity of hypercomplex orbitals, the resulting density and kinetic energy remain physically meaningful and satisfy essential properties such as the Pauli exclusion principle.

To further enrich HCOFT, an alternative definition of hypercomplex orbitals is explored, expanding the basis by incorporating new types of basis elements from Clifford algebra. This extension introduces elements that square to 1 and 0, in addition to the original elements that square to -1, providing greater flexibility in representing correlated electronic states and potentially enabling the treatment of a broader range of systems. Despite the increased flexibility, the derivations confirm that key physical quantities--such as density, spin density, and kinetic energy--remain real-valued and can be expressed in forms similar to those in the original HCOFT. However, a challenge regarding the positivity of the matrix $D^\sigma$ , used to compute the HCOs and their occupations, has been identified, as positive definiteness cannot be guaranteed. As a result, the potential application of the newly defined hypercomplex orbitals to many-electron systems remains an area for further theoretical development. This underscores the distinctive nature of hypercomplex orbitals in the original HCOFT, which not only preserves the desired physical properties but also allows a single-determinant framework to describe multi-reference systems.

%
%
\begin{acknowledgments}
Support from the National Natural Science Foundation of China (Grants No. 22473061, No. 22122303, and No. 22073049)  and Fundamental Research Funds for the Central Universities (Nankai University, Grant No. 63206008) is appreciated. 
\end{acknowledgments}

\bibliographystyle{aip}
\bibliography{ref}

\begin{thebibliography}{10}

\bibitem{HK1964}
P.~Hohenberg and W.~Kohn,
\newblock Phys. Rev. {\bf 136}, B864 (1964).

\bibitem{Levy1979pnas}
M.~Levy,
\newblock Proc. Natl. Acad. Sci. USA {\bf 76}, 6062 (1979).

\bibitem{KS1965}
W.~Kohn and L.~J. Sham,
\newblock Phys. Rev. {\bf 140}, A1133 (1965).

\bibitem{PY1989}
R.~G. Parr and W.~Yang,
\newblock {\em Density-Functional Theory of Atoms and Molecules},
\newblock Oxford University Press: New York, 1989.

\bibitem{Dreizler2012}
R.~Dreizler and E.~Gross,
\newblock {\em Density Functional Theory: An Approach to the Quantum Many-Body
  Problem},
\newblock Springer Berlin Heidelberg, 2012.

\bibitem{Cohen2008science}
A.~J. Cohen, P.~Mori-S\'anchez, and W.~Yang,
\newblock Science {\bf 321}, 792 (2008).

\bibitem{Paula2009prl}
P.~Mori-S\'anchez, A.~J. Cohen, and W.~Yang,
\newblock Phys. Rev. Lett. {\bf 102}, 066403 (2009).

\bibitem{Burke2012jcp}
K.~Burke,
\newblock J. Chem. Phys. {\bf 136}, 150901 (2012).

\bibitem{Cohen2012cr}
A.~J. Cohen, P.~Mori-S\'anchez, and W.~Yang,
\newblock Chem. Rev. {\bf 112}, 289 (2012).

\bibitem{Becke2014jcp}
A.~D. Becke,
\newblock J. Chem. Phys. {\bf 140}, 18A301 (2014).

\bibitem{Mardirossian2017}
N.~Mardirossian and M.~Head-Gordon,
\newblock Mol. Phys. {\bf 115}, 2315 (2017).

\bibitem{Xu2004}
X.~Xu and W.~A. Goddard,
\newblock J. Chem. Phys. {\bf 121}, 4068 (2004).

\bibitem{DFT2022PCCP}
A.~M. Teale et~al.,
\newblock Phys. Chem. Chem. Phys. {\bf 24}, 28700 (2022).

\bibitem{Gilbert1975prb}
T.~L. Gilbert,
\newblock Phys. Rev. B {\bf 12}, 2111 (1975).

\bibitem{Valone1980jcp}
S.~M. Valone,
\newblock J. Chem. Phys. {\bf 73}, 4653 (1980).

\bibitem{Muller1984rpa}
A.~Muller,
\newblock Phys. Lett. A {\bf 105}, 446  (1984).

\bibitem{GU1998prl}
S.~Goedecker and C.~J. Umrigar,
\newblock Phys. Rev. Lett. {\bf 81}, 866 (1998).

\bibitem{Pernal2005prl}
K.~Pernal,
\newblock Phys. Rev. Lett. {\bf 94}, 233002 (2005).

\bibitem{Sharma2008prb}
S.~Sharma, J.~K. Dewhurst, N.~N. Lathiotakis, and E.~K.~U. Gross,
\newblock Phys. Rev. B {\bf 78}, 201103(R) (2008).

\bibitem{Piris2010jcp}
M.~Piris, J.~M. Matxain, X.~Lopez, and J.~M. Ugalde,
\newblock J. Chem. Phys. {\bf 132}, 031103 (2010).

\bibitem{Sharma2013prl}
S.~Sharma, J.~K. Dewhurst, S.~Shallcross, and E.~K.~U. Gross,
\newblock Phys. Rev. Lett. {\bf 110}, 116403 (2013).

\bibitem{Schade2017}
R.~{Schade}, E.~{Kamil}, and P.~{Bl{\"o}chl},
\newblock Eur. Phys. J. Special Topics {\bf 226}, 2677 (2017).

\bibitem{Christian2018}
C.~Schilling,
\newblock J. Chem. Phys. {\bf 149}, 231102 (2018).

\bibitem{Schilling2019prl}
C.~Schilling and R.~Schilling,
\newblock Phys. Rev. Lett. {\bf 122}, 013001 (2019).

\bibitem{Cioslowski2020jctc}
J.~Cioslowski,
\newblock J. Chem. Theory Comput. {\bf 16}, 1578 (2020).

\bibitem{Cioslowski2020jcp}
J.~Cioslowski,
\newblock J. Chem. Phys. {\bf 153}, 154108 (2020).

\bibitem{Yao2021jpcl}
Y.-F. Yao, W.-H. Fang, and N.~Q. Su,
\newblock J. Phys. Chem. Lett. {\bf 12}, 6788 (2021).

\bibitem{Piris2021prl}
M.~Piris,
\newblock Phys. Rev. Lett. {\bf 127}, 233001 (2021).

\bibitem{Gibney2021jpcl}
D.~Gibney, J.-N. Boyn, and D.~A. Mazziotti,
\newblock J. Phys. Chem. Lett. {\bf 12}, 385 (2021).

\bibitem{Yao2022jpca}
Y.-F. Yao, Z.~Zhang, W.-H. Fang, and N.~Q. Su,
\newblock J. Phys. Chem. A {\bf 126}, 5654 (2022).

\bibitem{Gibney2022jctc}
D.~Gibney, J.-N. Boyn, and D.~A. Mazziotti,
\newblock J. Chem. Theory Comput. {\bf 18}, 6600 (2022).

\bibitem{Ai2022jpcl}
W.~Ai, W.-H. Fang, and N.~Q. Su,
\newblock J. Phys. Chem. Lett. {\bf 13}, 1744 (2022).

\bibitem{Ai2023jcp}
W.~Ai, N.~Q. Su, and W.-H. Fang,
\newblock J. Chem. Phys. {\bf 159}, 174110 (2023).

\bibitem{Gibney2023prl}
D.~Gibney, J.-N. Boyn, and D.~A. Mazziotti,
\newblock Phys. Rev. Lett. {\bf 131}, 243003 (2023).

\bibitem{Liebert2023jcp}
J.~Liebert, A.~Y. Chaou, and C.~Schilling,
\newblock J. Chem. Phys. {\bf 158}, 214108 (2023).

\bibitem{Yao2024jpca}
Y.-F. Yao and N.~Q. Su,
\newblock J. Phys. Chem. A {\bf 128}, 7669 (2024).

\bibitem{Cartier2024jctc}
N.~G. Cartier and K.~J.~H. Giesbertz,
\newblock J. Chem. Theory Comput. {\bf 20}, 3669 (2024).

\bibitem{Cartier2025arXiv}
N.~Cartier and K.~Giesbertz,
\newblock Impact of parametrizations of the one-body reduced density matrix on
  the energy landscape, 2025,
\newblock arXiv:2501.18996.

\bibitem{Voutou2025arXiv}
\'{E}lodie Boutou, J.~F.~H. Lew-Yee, J.~M. Mercero, and M.~Piris,
\newblock Enhancing the computational efficiency of the donof program through a
  new orbital sorting scheme, 2025,
\newblock arXiv:2502.01786.

\bibitem{Su2021pra}
N.~Q. Su,
\newblock Phys. Rev. A {\bf 104}, 052809 (2021).

\bibitem{Su2024prl}
N.~Q. Su,
\newblock Phys. Rev. Lett. {\bf 133}, 206402 (2024).

\bibitem{Szabo1996modern}
A.~Szabo and N.~Ostlund,
\newblock {\em Modern Quantum Chemistry: Introduction to Advanced Electronic
  Structure Theory},
\newblock Dover Publications, 1996.

\bibitem{HC1989}
I.~L. Kantor and A.~S. Solodovnikov,
\newblock {\em Hypercomplex numbers},
\newblock Springer-Verlag: Berlin, 1989.

\bibitem{CA2019}
J.~{Vaz Jr.} and R.~da~{Rocha Jr.},
\newblock {\em An Introduction to Clifford Algebras and Spinors},
\newblock Oxford University Press: New York, 2019.

\bibitem{Cayley1845}
A.~Cayley,
\newblock Philos. Mag. {\bf 26}, 141 (1845).

\bibitem{Study1920}
E.~Study,
\newblock Acta Math. {\bf 42}, 1 (1920).

\bibitem{Moore1922}
E.~H. Moore,
\newblock Bull. Amer. Math. Soc. {\bf 28}, 161 (1922).

\bibitem{Chen1991}
L.~Chen,
\newblock Acta Math. Sinica {\bf 7}, 171 (1991).

\bibitem{Kyrchei2008}
I.~I. Kyrchei,
\newblock J. Math. Sci. {\bf 155}, 839 (2008).

\bibitem{Aslaksen1996}
H.~Aslaksen,
\newblock Math. Intell. {\bf 18}, 57 (1996).

\bibitem{Lounesto2001}
P.~Lounesto,
\newblock {\em Clifford Algebras and Spinors},
\newblock Cambridge University Press, 2001.

\bibitem{Su2022es}
N.~Q. Su,
\newblock Electron. Struct. {\bf 4}, 014011 (2022).

\bibitem{Zhang2023pra}
T.~Zhang and N.~Q. Su,
\newblock Phys. Rev. A {\bf 108}, 052801 (2023).

\bibitem{Selig2005}
J.~M. Selig,
\newblock {\em Geometric Fundamentals of Robotics},
\newblock Springer New York, 2005.

\end{thebibliography}

\end{document}